# IMPROVING PARTITION-BLOCK-BASED ACOUSTIC ECHO CANCELER IN UNDER-MODELING SCENARIOS

*Wenzhi Fan*[1,2,3], *Jing Lu*[1,2]

[1] Key Laboratory of Modern Acoustics and Institute of Acoustics, Nanjing University, Nanjing, China, 210093
[2] NJU-Horizon Intelligent Audio Lab, Nanjing Institute of Advanced Artificial Intelligence, Nanjing, China, 210014
[3] ByteDance AI Lab
wenzhi_fan@smail.nju.edu.cn, lujing@nju.edu.cn

## Abstract

Recently, a partitioned-block-based frequency-domain Kalman filter (PFKF) has been proposed for acoustic echo cancellation. Compared with the normal frequency-domain Kalman filter, the PFKF utilizes the partitioned-block structure, resulting in both fast convergence and low time-latency. We present an analysis of the steady-state behavior of the PFKF and found that it suffers from a biased steady-state solution when the filter is of deficient length. Accordingly, we propose an effective modification that has the benefit of the guaranteed optimal steady-state behavior. Simulations are conducted to validate the improved performance of the proposed method.

**Index Terms**: Adaptive filter, Kalman filter, Acoustic echo cancellation

## 1. Introduction

Acoustic echo cancellation (AEC) has been one of the most challenging system identification problems in several decades. Due to acoustic coupling between the loudspeaker and the microphone of a telecommunication terminal, the far-end talker receives a delayed version of his own voice, which degrades the speech intelligibility and listening comfort [1]. The acoustic echo also seriously deteriorates the performance of the intelligent speech interaction system like the smart speaker. The adaptive algorithms, such as the least mean square (LMS), the recursive least squares (RLS), and their variant frequency-domain and sub-band implementations, are commonly applied to estimate the echo in the microphone signal [2]-[9]. A stochastic state-space model of the acoustic echo path has been derived by G. Enzner [10] and the corresponding frequency-domain Kalman filter (FKF) for acoustic echo cancellation is proposed. Subsequently, the FKF and its variations have been further developed and implemented in acoustic echo cancellation and many other fields [11]-[15].

Compared with the normally used frequency-domain adaptive filters, the FKF is a good tradeoff among the convergence rate, misalignment, robustness to double-talk and tracking of the typical echo-path changes [10][13]. Moreover, it does not require additional regularization or control mechanisms and is computationally efficient [10]. More recently, the FKF has been extended to the partitioned-block filtering structure [17] to ensure both fast convergence and low algorithm latency, which is of practical importance for AEC applications. It is generally assumed that the adaptive filter is of sufficient length in literatures for both the FKF and its partitioned-block-based variation (PFKF) [10]-[17]. However, in practical AEC applications, the impulse response of the echo path can be extremely long [18]-[20], resulting in under-modeling situations. Therefore, it is meaningful to investigate the performance of the PFKF when the filter is of deficient length.

In this paper, the steady-state behavior of the PFKF is analyzed by investigating the optimal solution of the equivalent weight vector in time-domain. It is found that the PFKF converges to a biased steady-state solution when the filter is of deficient length and the performance might deteriorate considerably. The normal frequency-domain Kalman filter and other frequency-domain adaptive filters also suffer from similar problems [20]-[23]. To resolve the problem of performance deterioration, a modification of the PFKF is proposed on the basis of the analysis, leading to a guaranteed optimal steady-state behavior. The modified version can be seen as an extension of [23] with partitioned-block structure. Numerical simulations are carried out to verify its performance.

## 2. Analysis of The Steady-state Behavior

The structure of the PFKF is briefly revisited as follows [17]. Let $y(n)$ be the microphone signal and $s(n)$ be the observation noise (the near-end signal in AEC). Since the echo signal results from the convolution of the reference signal $x(n)$ with the acoustic echo path $w(n)$, we have

$$y(n) = w(n) * x(n) + s(n), \quad (1)$$

where * denotes convolution. Assuming the length of the echo path is $N$ and dividing the filter with coefficients $w(n)$ into $B$ partitions of length $L$, the $b$-th partition of the filter coefficient vector at frame index $k$ can be described as

$$\mathbf{w}_b(k) = \left[w(bL,k), w(bL+1,k),...,w(bL+L-1,k)\right]^\mathrm{T}. \quad (2)$$

The time-domain reference signal vector for the $b$-th partition of length $M=2L$ with a frameshift of $L$ can be described as

$$\mathbf{x}_b(k) = \left[x(kL-bL-M+1),...,x(kL-bL)\right]^\mathrm{T}. \quad (3)$$

Then, the reference matrix in the frequency-domain can be denoted as

$$\mathbf{X}_b(k) = \mathrm{diag}\{\mathbf{F}\mathbf{x}_b(k)\}, \tag{4}$$

where $\mathbf{F}$ is an $M \times M$ Fourier transform matrix and $\mathrm{diag}\{\cdot\}$ creates a diagonal matrix from its input. Applying overlap-save method to compute a block of microphone signal, the filter output in frequency-domain can be denoted as

$$\mathbf{Y}(k) = \mathbf{G}_{0,L} \sum_{b=0}^{B-1} \mathbf{X}_b(k)\mathbf{W}_b(k) + \mathbf{S}(k), \tag{5}$$

where $\mathbf{Y}(k) = \mathbf{F}[\mathbf{0}_{1 \times L}, \mathbf{y}^T(k)]^T$ is the frequency-domain microphone signal vector, $\mathbf{S}(k) = \mathbf{F}[\mathbf{0}_{1 \times L}, \mathbf{s}^T(k)]^T$ is the frequency-domain near-end signal vector, and $\mathbf{W}_b(k) = \mathbf{F}[\mathbf{w}_b^T(k), \mathbf{0}_{1 \times L}]^T$ is the frequency-domain coefficients for the $b$-th partition. $\mathbf{G}_{0,L}$ is a constraining matrix as

$$\mathbf{G}_{0,L} = \mathbf{F}\begin{bmatrix} \mathbf{0}_L & \mathbf{0}_L \\ \mathbf{0}_L & \mathbf{I}_L \end{bmatrix}\mathbf{F}^{-1}, \tag{6}$$

with $\mathbf{I}_L$ a $L \times L$ identity matrix. The signal vectors $\mathbf{y}(k) = [y(kL-L+1), \ldots, y(kL)]^T$ and $\mathbf{s}(k) = [s(kL-L+1), \ldots, s(kL)]^T$ represent the $L$ latest samples of the microphone signal and the observation noise respectively.

A first order statistical Markov model is used to describe the time-varying property of the filter coefficients for the $b$-th partition [16] as

$$\mathbf{W}_b(k+1) = A \cdot \mathbf{W}_b(k) + \Delta \mathbf{W}_b(k), \tag{7}$$

where $A$ is the transition parameter and $\Delta\mathbf{W}_b(k)$ is the process noise vector. Analogous to [10], the desired stochastic state-space model for the filter partitions is then defined by the Markov model in (7) together with the linear observation model in (5).

In order to decrease the computational complexity, a simplified version of the PFKF was proposed in [17] as follows,

$$\hat{\mathbf{W}}_b(k+1) = A\left[\hat{\mathbf{W}}_b(k) + \mathbf{G}_{L,0}\boldsymbol{\mu}_b(k)\mathbf{X}_b(k)\mathbf{E}(k)\right], \tag{8}$$

$$\mathbf{K}_b(k+1) = \boldsymbol{\mu}_b(k)\mathbf{X}_b^H(k), \tag{9}$$

$$\boldsymbol{\mu}_b(k+1) = \frac{L}{M}\mathbf{P}_b(k)\left[\sum_{b=0}^{B-1}\mathbf{X}_b(k)\mathbf{P}_b(k)\mathbf{X}_b(k) + \boldsymbol{\Psi}_{SS}(k)\right]^{-1}, \tag{10}$$

$$\mathbf{P}_b(k+1) = A^2\left(\mathbf{I}_M - \frac{L}{M}\cdot\boldsymbol{\mu}_b(k)\mathbf{X}_b^H(k)\mathbf{X}_b(k)\right)\mathbf{P}_b(k) + \boldsymbol{\Psi}_{b,\Delta}(k), \tag{11}$$

$$\mathbf{E}(k) = \mathbf{Y}(k) - \mathbf{G}_{0,L}\sum_{b=0}^{B-1}\mathbf{X}_b(k)\hat{\mathbf{W}}_b(k), \tag{12}$$

where $\hat{\mathbf{W}}_b(k) = [\hat{W}_{b,0}(k), \ldots, \hat{W}_{b,M-1}(k)]^T = \mathbf{F}[\hat{\mathbf{w}}_b^T(k), \mathbf{0}_{1 \times L}]^T$ and $\mathbf{E}(k)$ are the frequency-domain system estimate vector and error vector respectively, $\hat{\mathbf{w}}_b(k)$ is the time-domain estimate vector, $\boldsymbol{\mu}_b(k) = \mathrm{diag}\{[\mu_{b,0}(k),\ldots,\mu_{b,M-1}(k)]^T\}$ is the step-size matrix, $\mathbf{K}_b(k)$ is the Kalman gain, the superscript H represents the conjugate transpose operation, $\boldsymbol{\Psi}_{b,\Delta}(k)$ and $\boldsymbol{\Psi}_{SS}(k)$ are the covariance matrix of the process noise and observation noise respectively, $\mathbf{P}_b(k) = \mathrm{diag}\{[P_{b,0}(k), \ldots, P_{b,M-1}(k)]^T\}$ is the state estimation error covariance matrix based on the Kalman filter theory [24], and $\mathbf{G}_{L,0}$ is another constraining matrix as

$$\mathbf{G}_{L,0} = \mathbf{F}\begin{bmatrix} \mathbf{I}_L & \mathbf{0}_L \\ \mathbf{0}_L & \mathbf{0}_L \end{bmatrix}\mathbf{F}^{-1}. \tag{13}$$

To analyze the PFKF, multiply both sides of (8) by $\mathbf{F}^{-1}$ and setting $A=1$ and obtain

$$\begin{bmatrix} \hat{\mathbf{w}}_b(k+1) \\ \mathbf{0}_{L \times 1} \end{bmatrix} = \begin{bmatrix} \hat{\mathbf{w}}_b(k) \\ \mathbf{0}_{L \times 1} \end{bmatrix} + \begin{bmatrix} \mathbf{I}_L & \mathbf{0}_L \\ \mathbf{0}_L & \mathbf{0}_L \end{bmatrix} \mathbf{M}_b(k)\mathbf{X}_{C,b}(k)\begin{bmatrix} \mathbf{0}_{L \times 1} \\ \mathbf{e}(k) \end{bmatrix}, \tag{14}$$

where $\mathbf{e}(k) = [e(kL-L+1), \ldots, e(kL)]^T$,

$$\mathbf{X}_{C,b}(k) = \mathbf{F}^{-1}\mathbf{X}_b(k)\mathbf{F} = \begin{bmatrix} \mathbf{X}_{C,b,1}(k) & \mathbf{X}_{C,b,2}(k) \\ \mathbf{X}_{C,b,2}(k) & \mathbf{X}_{C,b,1}(k) \end{bmatrix} \tag{15}$$

is a circulant matrix whose first row is $\mathbf{x}_b(k)$ and

$$\mathbf{M}_b(k) = \mathbf{F}^{-1}\boldsymbol{\mu}_b(k)\mathbf{F} = \begin{bmatrix} \mathbf{M}_{b,1}(k) & \mathbf{M}_{b,2}(k) \\ \mathbf{M}_{b,2}(k) & \mathbf{M}_{b,1}(k) \end{bmatrix} \tag{16}$$

is also a circulant matrix whose first row is $\mathbf{F}^{-1}[\mu_{b,0}(k), \mu_{b,1}(k), \ldots, \mu_{b,M-1}(k)]^T$. $\mathbf{X}_{C,b,1}(k)$, $\mathbf{X}_{C,b,2}(k)$, $\mathbf{M}_{b,1}(k)$ and $\mathbf{M}_{b,2}(k)$ are matrices with size $L \times L$. Substitute (15) and (16) into (14), the time-domain update equation for the PFKF can be described as

$$\hat{\mathbf{w}}_b(k+1) = \hat{\mathbf{w}}_b(k) + \left[\mathbf{M}_{b,1}(k)\mathbf{X}_{C,b,2}(k) + \mathbf{M}_{b,2}(k)\mathbf{X}_{C,b,1}(k)\right]\mathbf{e}(k). \tag{17}$$

The time-domain error vector can be rearranged as

$$\mathbf{e}(k) = \mathbf{y}(k) - \sum_{b=0}^{B-1}\mathbf{X}_{C,b,2}^T(k)\hat{\mathbf{w}}_b(k). \tag{18}$$

To analyze the convergence behavior of the system, the reference signal and the filter coefficients are regarded independent, which is a common assumption in adaptive filter analysis [2]. Additionally, the step-size matrix $\boldsymbol{\mu}_b(k)$, as well as its related matrix $\mathbf{M}_b(k)$, is assumed to be independent of the reference signal and the filter coefficients, which is widely assumed in the analysis of variable step-size adaptive algorithm [25]-[27]. The mean convergence behavior of the time-domain filter coefficients can be determined by taking expectation on both sides of (17) as

$$\mathrm{E}\{\hat{\mathbf{w}}_b(k+1)\} = \mathrm{E}\{\hat{\mathbf{w}}_b(k)\} + \left[\boldsymbol{\Lambda}_{b,1}(k)\mathbf{r}_b + \boldsymbol{\Lambda}_{b,2}(k)\hat{\mathbf{r}}_b\right] - \left[\boldsymbol{\Lambda}_{b,1}(k)\sum_{m=0}^{B-1}\mathbf{R}_{b,m}\mathrm{E}\{\hat{\mathbf{w}}_m(k)\} + \boldsymbol{\Lambda}_{b,2}(k)\sum_{m=0}^{B-1}\hat{\mathbf{R}}_{b,m}\mathrm{E}\{\hat{\mathbf{w}}_m(k)\}\right], \tag{19}$$

where

$$\mathbf{R}_{b,m} = \mathrm{E}\{\mathbf{X}_{C,b,2}(k)\mathbf{X}_{C,m,2}^T(k)\}, \quad \hat{\mathbf{R}}_{b,m} = \mathrm{E}\{\mathbf{X}_{C,b,1}(k)\mathbf{X}_{C,m,2}^T(k)\},$$
$$\mathbf{r}_b = \mathrm{E}\{\mathbf{X}_{C,b,2}(k)\mathbf{y}(k)\}, \quad \hat{\mathbf{r}}_b = \mathrm{E}\{\mathbf{X}_{C,b,1}(k)\mathbf{y}(k)\},$$
$$\boldsymbol{\Lambda}_{b,1}(k) = \mathrm{E}\{\mathbf{M}_{b,1}(k)\}, \quad \boldsymbol{\Lambda}_{b,2}(k) = \mathrm{E}\{\mathbf{M}_{b,2}(k)\}. \tag{20}$$

It is found (19) can be written in a compact form as

$$\mathrm{E}\{\underline{\hat{\mathbf{w}}}(k+1)\} = \left[\mathbf{I}_M - \underline{\boldsymbol{\Lambda}}_1(k)\underline{\mathbf{R}} - \underline{\boldsymbol{\Lambda}}_2(k)\underline{\hat{\mathbf{R}}}\right]\mathrm{E}\{\underline{\hat{\mathbf{w}}}(k)\} + \left[\underline{\boldsymbol{\Lambda}}_1(k)\underline{\mathbf{r}} + \underline{\boldsymbol{\Lambda}}_2(k)\underline{\hat{\mathbf{r}}}\right], \tag{21}$$

by defining the super vectors as $\underline{\hat{\mathbf{w}}}(k)=[\hat{\mathbf{w}}^T_0(k), \ldots, \hat{\mathbf{w}}^T_{B-1}(k)]^T$, $\underline{\mathbf{r}}=[\mathbf{r}^T_0,\ldots,\mathbf{r}^T_{B-1}]^T$, $\underline{\hat{\mathbf{r}}}=[\hat{\mathbf{r}}^T_0,\ldots,\hat{\mathbf{r}}^T_{B-1}]^T$, and the super matrices as

$$\underline{\mathbf{R}}=\begin{bmatrix} \mathbf{R}_{0,0} & \cdots & \mathbf{R}_{0,B-1} \\ \vdots & & \vdots \\ \mathbf{R}_{B-1,0} & \cdots & \mathbf{R}_{B-1,B-1} \end{bmatrix}, \underline{\hat{\mathbf{R}}}=\begin{bmatrix} \hat{\mathbf{R}}_{0,0} & \cdots & \hat{\mathbf{R}}_{0,B-1} \\ \vdots & & \vdots \\ \hat{\mathbf{R}}_{B-1,0} & \cdots & \hat{\mathbf{R}}_{B-1,B-1} \end{bmatrix}, \quad (22)$$

$$\underline{\mathbf{\Lambda}}_i(k)=\begin{bmatrix} \mathbf{\Lambda}_{0,i}(k) & \cdots & \mathbf{0}_L \\ \vdots & & \vdots \\ \mathbf{0}_L & \cdots & \mathbf{\Lambda}_{B-1,i}(k) \end{bmatrix}, i=1,2. \quad (23)$$

The steady-state solution of (21) can be obtained as

$$E\{\underline{\hat{\mathbf{w}}}(\infty)\} = \left[\underline{\mathbf{\Lambda}}_1(\infty)\underline{\mathbf{R}} + \underline{\mathbf{\Lambda}}_2(\infty)\underline{\hat{\mathbf{R}}}\right]^{-1} \cdot \left[\underline{\mathbf{\Lambda}}_1(\infty)\underline{\mathbf{r}} + \underline{\mathbf{\Lambda}}_2(\infty)\underline{\hat{\mathbf{r}}}\right]. \quad (24)$$

For the situation of a sufficient filter length, i.e. the length of the actual acoustic echo path $N'\leq N$, the echo path can be described as $\mathbf{w}_o = [w_{o,0},\ldots,w_{o,N'-1},\mathbf{0}_{1\times(N-N')}]^T$. Then, the microphone signal can be denoted as

$$\mathbf{y}(k) = \mathbf{s}(k) + \sum_{m=0}^{B-1} \mathbf{X}^T_{C,m,2}(k)\mathbf{w}_{o,m}. \quad (25)$$

with $\mathbf{w}_{o,m}=[w_{o,mL},\ldots,w_{o,mL+L-1}]^T$. Assuming that the observation noise is independent from the reference signal and substituting (25) into (20) yields

$$\underline{\mathbf{r}} = \underline{\mathbf{R}}\mathbf{w}_o, \underline{\hat{\mathbf{r}}} = \underline{\hat{\mathbf{R}}}\mathbf{w}_o. \quad (26)$$

Combining (24) and (26), it can be seen from that the solution for $E\{\underline{\hat{\mathbf{w}}}(\infty)\}$ is exactly $\mathbf{w}_o$, implying the optimality of the PFKF when the filter length is sufficient. When the filter is of deficient length, (25) and (26) cannot be attained, a biased steady-state solution of the PFKF is thus unavoidable.

## 3. The Proposed Method

With careful observation on (17) and (24), it is found that the term, $\mathbf{M}_{b,2}(k)\mathbf{X}_{C,b,1}(k)$, obstructs the PFKF converging to the optimal steady-state solution. The update equation (8) can be revised by changing the position of the constraining matrix $\mathbf{G}_{L,0}$ to circumvent the unfavorable effect of $\mathbf{M}_{b,2}(k)\mathbf{X}_{C,b,1}(k)$:

$$\hat{\mathbf{W}}_b(k+1) = A\left[\hat{\mathbf{W}}_b(k) + \boldsymbol{\mu}_b(k)\mathbf{G}_{L,0}\mathbf{X}^H_b(k)\mathbf{E}(k)\right]. \quad (27)$$

Multiplying both sides of by $\mathbf{F}^{-1}$ and setting $A=1$ yields

$$\begin{bmatrix} \hat{\mathbf{w}}_b(k+1) \\ \hat{\mathbf{w}}_{wr,b}(k+1) \end{bmatrix} = \begin{bmatrix} \hat{\mathbf{w}}_b(k) \\ \hat{\mathbf{w}}_{wr,b}(k) \end{bmatrix} + \mathbf{M}_b(k)\begin{bmatrix} \mathbf{I}_L & \mathbf{0}_L \\ \mathbf{0}_L & \mathbf{0}_L \end{bmatrix}\mathbf{X}_{C,b}(k)\begin{bmatrix} \mathbf{0}_{L\times 1} \\ \mathbf{e}(k) \end{bmatrix}, \quad (28)$$

where $\hat{\mathbf{w}}_{wr,b}(k)$ represents the part of filter coefficients that suffers from the wraparound effect of circular convolution. Focusing on the causal part of the filter coefficients, the following update equation in time-domain can be attained:

$$\hat{\mathbf{w}}_b(k+1) = \hat{\mathbf{w}}_b(k) + \mathbf{M}_{b,1}(k)\mathbf{X}_{C,b,2}(k)\mathbf{e}(k). \quad (29)$$

Taking expectation on both sides of (29) yields

$$E\{\hat{\mathbf{w}}_b(k+1)\} = E\{\hat{\mathbf{w}}_b(k)\} - \mathbf{\Lambda}_{b,1}(k)\sum_{m=0}^{B-1}\mathbf{R}_{b,m}E\{\hat{\mathbf{w}}_m(k)\} + \mathbf{\Lambda}_{b,1}(k)\mathbf{r}. \quad (30)$$

Similarly, (30) can be written in a compact form as

$$E\{\underline{\hat{\mathbf{w}}}(k+1)\} = \left[\mathbf{I}_M - \underline{\mathbf{\Lambda}}_1(k)\underline{\mathbf{R}}\right] \cdot E\{\underline{\hat{\mathbf{w}}}(k)\} + \underline{\mathbf{\Lambda}}_1(k)\underline{\mathbf{r}}, \quad (31)$$

and its steady-state solution can be obtained as $E\{\underline{\hat{\mathbf{w}}}(\infty)\}=\underline{\mathbf{R}}^{-1}\cdot\underline{\mathbf{r}}$. It can found from (20) that

$$\mathbf{R}_{i,j} = E\{\mathbf{X}_{C,0,2}(k-i)\mathbf{X}^T_{C,0,2}(k-j)\} = L\mathbf{R}_{x,i,j}, \quad (32)$$
$$\mathbf{r}_b = E\{\mathbf{X}_{C,0,2}(k-b)\mathbf{y}(k)\} = L\mathbf{p}_b,$$

where $\mathbf{p}_i=[p_{iL},\ldots,p_{iL+L-1}]^T$ and

$$\mathbf{R}_{x,i,j} = \begin{bmatrix} R_x(|(i-j)L|) & \cdots & R_x(|(i-j)L-L+1|) \\ \vdots & & \vdots \\ R_x(|(i-j)L+L-1|) & \cdots & R_x(|(i-j)L|) \end{bmatrix}. \quad (33)$$

Here, $R_x(i)=E\{x(n)x(n-i)\}$ is the auto-correlation of the reference signal with $R_x(i)=R_x(-i)$ and $p_i=E\{y(n)x(n-i)\}$ is the correlation between the desired signal and reference signal. Combining (22), (32) and (33), it can be found that $\underline{\mathbf{r}}=L\mathbf{p}$, $\underline{\mathbf{R}}=L\mathbf{R}_x$ and the steady-state solution of the modified method can be simplified as $E\{\underline{\hat{\mathbf{w}}}(\infty)\}=\mathbf{R}_x^{-1}\mathbf{p}$, where $\mathbf{R}_x$ is a $N\times N$ auto-correlation matrix of the reference signal and $\mathbf{p}=[\mathbf{p}^T_0,\ldots,\mathbf{p}^T_{B-1}]^T$ of length $N$ is the correlation vector between the reference signal and the microphone signal, implying the optimality of the modified solution in the sense of mean squared error.

It can be seen from (8) and (27) that the computational complexity of the original PFKF and the modified one is the same. However, extra constraints are needed to eliminate the wraparound effect of $\hat{\mathbf{w}}_{wr,b}(k)$ in the proposed algorithm when computing the output:

$$\hat{\mathbf{Y}}(k) = \mathbf{G}_{0,L}\sum_{b=0}^{B-1}\mathbf{X}_b(k)\mathbf{G}_{L,0}\hat{\mathbf{W}}_b(k). \quad (34)$$

Consequently, the modified algorithm achieves a guaranteed optimal solution with an extra computational load of $B$ pairs of $M$-point FFT/IFFT.

## 4. Computer Simulations

To demonstrate the convergence of algorithm, the normalized misalignment of the filter coefficients (in dB) is defined as

$$m(k) = 10\log_{10}\left[\frac{(\hat{\mathbf{w}}(k)-\hat{\mathbf{w}}_o)^T(\hat{\mathbf{w}}(k)-\hat{\mathbf{w}}_o)}{\|\hat{\mathbf{w}}_o\|^2}\right], \quad (35)$$

with $\hat{\mathbf{w}}_o$ the optimal solution calculated by $E\{\hat{\mathbf{w}}_o(\infty)\}=\mathbf{R}_x^{-1}\mathbf{p}$, which is similar to the definition in [28].

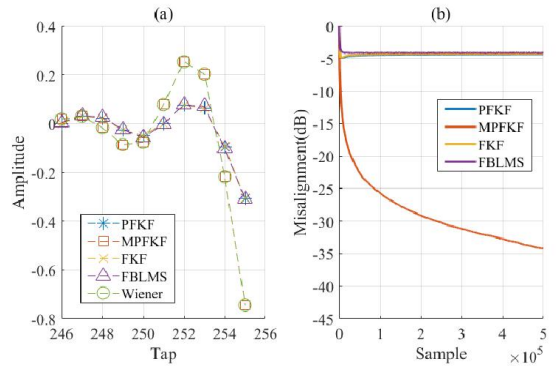

Figure 1: *(a) Steady-state solution of the last 10 taps of the filter coefficients. (b) Misalignments of the PFKF, FKF, MPFKF and FBLMS.*

### 4.1. An Illustrating Example

In this example, the reference signal is generated by passing Gaussian white noise with unit variance through a 4-tap lowpass FIR filter. The microphone signal is generated by passing the reference signal through a 512-tap high-pass FIR filter. An uncorrelated Gaussian white noise with SNR of 20 dB is added in the microphone signal. The power spectral density of the observation noise is assumed to be known. The length of the adaptive filter $N$ is 256, which is surely deficient to model the unknown system. The partition number $B$ is set to be 4, the block length $L$ and the frame-length $M$ are 64 and 128 accordingly. The transition parameter $A$ is set to be 1. The simulation results are averaged over 20 trials.

Fig. 1(a) depicts the last 10 taps of the time-domain steady-state filter coefficients in this under-modeling situation. Apparently, the steady-state solution of the proposed algorithm is in alignment with the optimal solution, whereas the FKF [10], PFKF and bin-normalized frequency-domain block LMS (FBLMS) [2] deviate from the Wiener solution. Fig. 1(b) depicts the misalignments of the four algorithms. It can be seen that the steady-state misalignment of the MPFKF is significantly smaller than other algorithms since its steady-state solution is unbiased. It is noted that the misalignment curves of the PFKF, FKF and FBLMS are flat since their fluctuations are masked by the comparatively large deviation of the steady-state solution in the logarithmic axis.

The misalignments of the PFKF and MPFKF with different settings of transition parameter in the same under-modeling example are shown in Fig.2. It has been addressed in [28] that this parameter has influence on the convergence rate, the tracking ability and the steady-state misalignment. Fig. 2 shows that the steady-state misalignment of the MPFKF decreases as the parameter $A$ increases and overall it is conspicuous that the MPFKF with different $A$ has an advantage over the standard PFKF in this situation.

### 4.2. A Practical AEC Example

The echo signal is simulated by convolving the reference signal (clean speech) with a measured room impulse response of 2048 taps with a reverberation time of about 1.2 seconds, as shown in Figure 3(a). An uncorrelated Gaussian white noise with SNR of 20 dB is added in the echo signal. The sampling rate is 16 kHz. The length of the adaptive filter $N$ is 1024 (64ms), which is significantly deficient for modeling the impulse response. The transition parameter $A$ is also set to be 1. Figure 3(b) depicts the misalignments of four algorithms in the actual AEC scenario. It can be seen that the modified algorithm has a lower misalignment than the FBLMS, FKF, PFKF and FBLMS as it approaches the steady-state, indicating a preferable performance in practical applications.

### 5. Conclusions

The steady-state behavior of the simplified partitioned-block-based frequency-domain Kalman filter has been investigated in this paper. It is found that the steady-state solution of the PFKF is biased in the under-modeling situation. On the basis of the analysis, a modification is proposed to improve the steady-state performance of the PFKF. The modified method achieves a guaranteed optimal steady-state solution. Numerical simulations validate the efficacy of the proposed algorithm for AEC applications.

### 6. Acknowledgements

This work was supported by the National Natural Science Foundation of China (Grant No. 11874219).

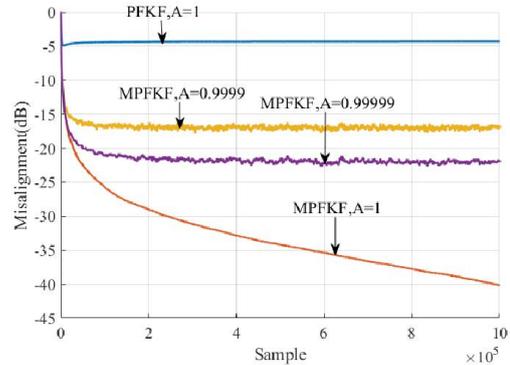

Figure 2: *Misalignments of PFKF and MPFKF with different transition parameter settings in the under-modeling example.*

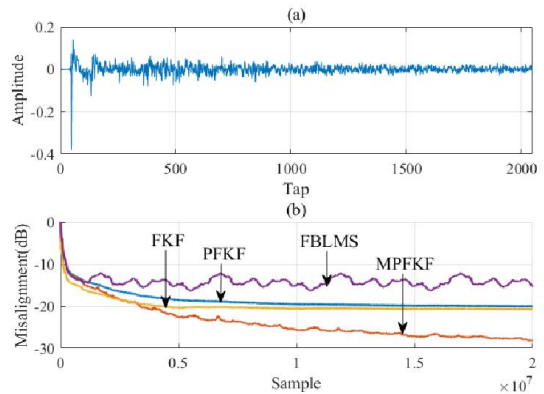

Figure 3: *(a)The impulse response of the acoustic echo path. (b)Misalignment curves of the PFKF and other algorithms in the AEC example*

### 7. References

[1] J. Benesty, M. M. Sondhi, and Y. Huang, *Springer handbook of speech processing*. Springer, 2007.

[2] B. Farhang-Boroujeny, in *Adaptive Filters: Theory and Applications*. Hoboken, NJ, USA: Wiley, 2013.

[3] J. Benesty, T. Gansler, D.R. Morgan, M. M. Sondhi, and S. L. Gay, Advances in Network and Acoustic Echo Cancellation, Springer, 2001.

[4] D. R. Morgan and J.C. Thi, "A delayless subband adaptive filter architecture," IEEE Trans. Signal Process., vol. 43, no. 8, pp. 1819–1830, August 1995.

[5] E. R. Ferrara, "Frequency-domain adaptive filtering," in Adaptive Filters, C.F.N. Cowan and P.M. Grant, Eds, pp. 145–179. Prentice Hall, Englewood Cliffs (NJ), 1985.

[6] J. J. Shynk, "Frequency-domain and multirate adaptive filtering," IEEE Signal Process. Mag., vol. 3, no. 2, pp. 14–37, Jan. 1992.

[7] K. S. Chan and B. Farhang-Boroujeny, "Analysis of the partitioned frequency-domain block LMS (PFBLMS) algorithm," *IEEE Transactions on Signal Processing,* vol. 49, no. 9, pp. 1860-1874, 2001.

[8] M. R. Asharif, T. Takebayashi, T. Chugo, and K. Murano, "Frequency domain noise canceler: Frequency-bin adaptive filtering (FBAF)," in Proc. ICASSP, 1986, pp. 41.22.1–41.22.4.

[9] E. Moulines, O. A. Amrane, and Y. Grenier, "The generalized multidelay adaptive filter: Structure and convergence analysis," IEEE Trans. Signal Processing, vol. 43, pp. 14–28, Jan. 1995.

[10] G. Enzner and P. Vary, "Frequency-domain adaptive Kalman filter for acoustic echo control in hands-free telephones," *Signal Processing,* vol. 86, no. 6, pp. 1140-1156, 2006.

[11] C. Wu, X. Wang, Y. Guo, Q. Fu, and Y. Yan, "Robust uncertainty control of the simplified Kalman filter for acoustic echo cancelation," Circuits Syst. Signal Process., vol. 35, no. 12, pp. 4584–4595, 2016.

[12] S. Malik and G. Enzner, "State-space frequency-domain adaptive filtering for nonlinear acoustic echo cancellation," IEEE Trans. Audio, Speech, Lang. Process., vol. 20, no. 7, pp. 2065–2079, Sep. 2012.

[13] F. Yang, G. Enzner, and J. Yang, "Frequency-domain adaptive Kalman filter with fast recovery of abrupt echo-path changes,"IEEE Signal Process. Lett., vol. 24, no. 12, pp. 1778–1782, Dec. 2017.

[14] B. Schwartz, S. Gannot, and E. A. P. Habets, "Online speech dereverberation using Kalman filter and EM algorithm,"IEEE Trans. Audio, Speech, Lang. Process., vol. 23, no. 2, pp. 394–406, Feb. 2015.

[15] G. Bernardi, T. V. Waterschoot, J. Wouters, M. Hillbmtt, and M. Moonen, "A PEM-based frequency-domain Kalman filter for adaptive feedback cancellation," inSignal Process. Conf., 2015, pp. 270–274.

[16] S. Haykin, *Adaptive Filter Theory*. Englewood Cliffs, NJ, USA: PrenticeHall, 2002.

[17] F. Kuech, E. Mabande, and G. Enzner, "State-space architecture of the partitioned-block-based acoustic echo controller," in *2014 IEEE International Conference on Acoustics, Speech and Signal Processing (ICASSP)*, 2014, pp. 1295-1299: IEEE.

[18] K. Mayyas, "Performance analysis of the deficient length LMS adaptive algorithm," *IEEE Transactions on Signal Processing,* vol. 53, no. 8, pp. 2727-2734, 2005.

[19] S. J. Elliott and B. Rafaely, "Frequency-domain adaptation of causal digital filters," *IEEE Transactions on Signal Processing,* vol. 48, no. 5, pp. 1354-1364, 2002.

[20] M. Wu, J. Yang, Y. Xu, and X. Qiu, "Steady-State Solution of the Deficient Length Constrained FBLMS Algorithm," *IEEE Transactions on Signal Processing,* vol. 60, no. 12, pp. 6681-6687, 2012.

[21] J. Lu, X. Qiu, and H. Zou, "A modified frequency-domain block LMS algorithm with guaranteed optimal steady-state performance," *Signal Processing,* vol. 104, no. 6, pp. 27-32, 2014.

[22] J. Lu, K. Chen, and X. Qiu, "Convergence analysis of the modified frequency-domain block LMS algorithm with guaranteed optimal steady state performance," *Signal Processing,* vol. 132, pp. 165-169, 2017.

[23] W. Fan, K. Chen, J. Lu, and J. Tao, "Effective improvement of under-modeling frequency-domain Kalman filter," *IEEE Signal Processing Letters,* 2019.

[24] B. D. O. Anderson, J. B. Moore, and M. Eslami, "Optimal Filtering," *IEEE Transactions on Systems Man & Cybernetics,* vol. 12, no. 2, pp. 235-236, 2007.

[25] H. S. Lee, S. E. Kim, J. W. Lee, and W. J. Song, "A Variable Step-Size Diffusion LMS Algorithm for Distributed Estimation," *IEEE Transactions on Signal Processing,* vol. 63, no. 7, pp. 1808-1820, 2015.

[26] K. Mayyas and F. Momani, "An LMS adaptive algorithm with a new step-size control equation," *Journal of the Franklin Institute,* vol. 348, no. 4, pp. 589-605, 2011.

[27] R. H. Kwong and E. W. Johnston, "A variable step size LMS algorithm," *IEEE Trans Signal Processing,* vol. 40, no. 7, pp. 1633-1642, 1992.

[28] F. Yang, G. Enzner, and J. Yang, "Frequency-Domain Adaptive Kalman Filter with Fast Recovery of Abrupt Echo-Path Changes," *IEEE Signal Processing Letters,* vol. PP, no. 99, pp. 1-1, 2017.